# Faraday patterns in low-dimensional Bose–Einstein condensates

Kestutis Staliunas[1,(1)], Stefano Longhi[2,(2)], and Germán J. de Valcárcel[3,(3)]

(1)*Physikalisch Technische Bundesanstalt, Bundesallee 100, D-38116 Braunschweig, Germany.*
(2)*INFM, Dipartimento di Fisica and IFN, Politecnico di Milano, Piazza L. da Vinci 32, I-20133 Milano, Italy.*
(3)*Departament d'Òptica, Universitat de València, Dr. Moliner 50, E-46100 Burjassot, Spain.*

**Abstract**

*We show that Faraday patterns can be excited in the weak confinement space of low-dimensional Bose-Einstein condensates by temporal modulation of the trap width, or equivalently of the trap frequency $\Omega_{tight}$, in the tight confinement space. For slow modulation, as compared with $\Omega_{tight}$, the low-dimensional dynamics of the condensate in the weak confinement space is described by a Gross-Pitaevskii equation with time modulated nonlinearity coefficient. For increasing modulation frequencies a noticeable reduction of the pattern formation threshold is observed close to $2\Omega_{tight}$, which is related to the parametric excitation of the internal breathing mode in the tight confinement space. These predictions could be relevant for the experimental excitation of Faraday patterns in Bose-Einstein condensates.*

---

[1] e-mail: kestutis.staliunas@ptb.de
[2] e-mail: longhi@fisi.polimi.it
[3] e-mail: german.valcarcel@uv.es

The nonlinear spatio-temporal dynamics of Bose-Einstein Condensates (BECs) is attracting great interest in the last few years. The interest concerns both spatially localized structures, like solitons and vortices, and spatially extended patterns. In particular the spontaneous emergence of spatially-extended patterns and quasipatterns has been predicted in BECs when the atomic scattering length is periodically modulated in time [1]. As these patterns arise due to the modulation of a system parameter (the scattering length), they show features similar to the parametric Faraday waves observed on the free surface of a fluid subjected to oscillatory vertical acceleration [2]. Indeed, the atomic density waves excited in this way oscillate at half the modulation frequency, and the selected wavenumber depends on the modulation frequency through a dispersion-induced mechanism. Time-periodic modulation of the scattering length can be achieved in practice by means of the so-called Feshbach resonance [3], which can be used as a tool for managing BECs for other purposes [4].

The main goal of this work is to show that, in low-dimensional BECs, Faraday patterns can be excited by the modulation of the trap width (equivalently the trap frequency) in the tight confinement direction, alternatively to the modulation of the atomic scattering length. In particular, two-dimensional (2D) Faraday patterns in disk-shaped BECs can be excited by a periodic modulation of the trap parameter in a direction normal to the disk plane, whereas one-dimensional (1D) Faraday waves can be excited in cigar-shaped BECs by periodic modulation of the radial confinement trap parameter. We also show that the mechanism introduces new resonance phenomena which manifest themselves, among others, by a noticeable lowering of the pattern formation threshold. These results are hence of interest for experimental observation of patterns in BECs, where modulation of trap frequencies is customary and usually easier than modulation of the atomic scattering length.

The starting point of our analysis is the Gross-Pitaevskii (GP) equation [5] for a confined BEC, generalized to include damping [6],

$$i\hbar \frac{\partial \Psi}{\partial t} = (1 - i\gamma)\left(-\frac{\hbar^2}{2m}\nabla^2 + V(\mathbf{r},t) + C|\Psi|^2 - \bar{\mu}\right)\Psi, \qquad (1)$$

where $C = 4\pi\hbar^2 Na/m$, $N$ is the number of particles, $a$ is the interatomic $s$-wave scattering length ($a > 0$ for a repulsive BEC, which we consider), $m$ is the mass of the particles, $\bar{\mu}$ is the chemical potential, $V(\mathbf{r},t)$ is the trapping potential, and $\gamma$ is the damping parameter. In the absence of damping the normalization condition is $\int |\Psi(\mathbf{r},t)|^2 d^3r = 1$. Our study covers both the 1D case of a cigar-shaped condensate extended along the $z$ direction, $V(\mathbf{r},t) = \frac{1}{2}m[\Omega_{tight}^2(t)(x^2+y^2) + \Omega_{weak}^2 z^2]$, and the 2D case of a disk-shaped condensate extended across the $(x,y)$ plane, $V(\mathbf{r},t) = \frac{1}{2}m[\Omega_{tight}^2(t)z^2 + \Omega_{weak}^2(x^2+y^2)]$. The condition $\Omega_{weak} \ll \Omega_{tight}$ is assumed which means that $\Omega_{weak}$ and $\Omega_{tight}$ are the frequencies of the trap along the weak and tight confinement directions respectively. We assume that the frequency $\Omega_{tight}$ is subjected to periodic modulation: $\Omega_{tight}(t) = \bar{\Omega}_{tight}[1 + \alpha\cos(\Omega t)]$. In terms of the scaled variables $\tau = \bar{\Omega}_{tight}t$, $\mathbf{R} \equiv (X,Y,Z) = (x,y,z)/a_{tight}$, ($a_{tight} = \sqrt{\hbar/(m\bar{\Omega}_{tight})}$ is the average width of the BEC along the tight confinement direction, in the noninteraction limit) and $u = a_{tight}\sqrt{4\pi a N}\,\Psi$, and assuming for definiteness a disk-shaped condensate, Eq. (1) takes the following dimensionless form:

$$i\frac{\partial u}{\partial \tau} = (1-i\gamma)\left[-\tfrac{1}{2}\nabla_{\mathbf{R}}^2 + \tfrac{1}{2}\omega_{tight}^2(\tau)Z^2 + \tfrac{1}{2}\omega_{weak}^2(X^2+Y^2) + |u|^2 - \mu\right]u, \qquad (2)$$

where

$$\omega_{tight}(\tau) = 1 + \alpha\cos(\omega\tau), \qquad (3)$$

$\omega_{weak} = \Omega_{weak}/\overline{\Omega}_{tight} \ll 1$, $\omega = \Omega/\overline{\Omega}_{tight}$, and $\mu = \overline{\mu}/(\hbar\overline{\Omega}_{tight})$. The dynamical equation for a cigar-shaped BEC is simply retrieved from Eq. (2) after interchanging $\omega_{weak}$ with $\omega_{tight}$. In the absence of damping, the normalization condition for $u$ reads $\int|u(\mathbf{R},\tau)|^2 d^3R = Q$, where $Q = 4\pi N a/a_{tight}$ is an adimensional parameter characterizing the strength of the nonlinear interaction. For modulation frequencies $\Omega$ small compared with $\Omega_{tight}$ ($0 < \omega \ll 1$), the dynamical equation (2) can be reduced to the same GP equation with time-modulated nonlinear coefficient, as used to describe Faraday patterns in BECs under scattering length modulation in Ref. [1]. The physical reason for the equivalence of modulating the scattering length and the trap frequency corresponding to the tightly confined direction ($z$) lies in the (periodically forced) breathing dynamics of the BEC across that direction. This dynamics entails a periodic change of the condensate density which, in its turn, leads to an effective modulation of the nonlinear particle interaction in the weakly confined ($x, y$) plane, hence becoming equivalent to the modulation of the scattering length. However, as $\Omega$ is increased, an additional resonance phenomenon is observed near $\Omega = 2\Omega_{tight}$ ($\omega \approx 2$), corresponding to the parametric excitation of the breathing mode of the condensate in the tight confinement direction [7]. Although close to resonance the dynamics becomes complicated and a rigorous reduction to a lower-dimensional GP equation seems not possible, pattern formation is still observed at much lower thresholds, which may be of interest from an experimental viewpoint.

We consider first the slow-modulation regime. Similarly to Ref. [8], the reduction of the BEC 3D dynamics to an effective 2D description is done through a multiple scale analysis. The derivation assumes a slow modulation frequency $\omega$, say of order of $\varepsilon$ (with $\varepsilon \ll 1$), a weak interaction of particles, $|u|^2 \propto O(\varepsilon)$, and a large (of order $\varepsilon^{-1/2}$) characteristic spatial scale of the condensate wave function variation across the weak confinement plane. This scaling corresponds to $Q \propto O(\varepsilon^0)$, which is compatible with typical experimental conditions, where $Q$ is in the range $10^1$–$10^3$ (see, e.g., [9]). Under these conditions a weakly nonlinear analysis of Eq. (2), that uses $\varepsilon^{1/2}$ as the expansion parameter, leads to the factorization

$$u(X,Y,Z,\tau) = \left[\tfrac{1}{2}\omega_{tight}(\tau)\right]^{1/4}\exp\left[-\tfrac{1}{2}\omega_{tight}(\tau)Z^2\right]\psi(X,Y,\tau), \qquad (4)$$

where, at the leading order, the reduced amplitude $\psi \propto O(\varepsilon^{1/2})$ satisfies

$$i\frac{\partial\psi}{\partial\tau} = \frac{1-i\gamma}{2}\left[-\nabla_{weak}^2 + \omega_{weak}^2(X^2+Y^2) + \sqrt{\omega_{tight}(\tau)}|\psi|^2 - \mu_1\right]\psi, \qquad (5)$$

which is obtained as a solvability condition at order $\varepsilon^{3/2}$ in the asymptotic expansion. The chemical potential is given by $\mu = [\omega_{tight}(\tau) + \mu_1]/2$, with $\mu_1 \propto O(\varepsilon)$ [10]. Note that the reduced equation is a damped GP equation with time-varying nonlinear term. This means that the modulation of the trap frequency corresponding to the tightly confined direction leads to an effective modulation of the particle interaction in the low-dimensional space. Assuming a flat trap in the weakly confined space ($\omega_{weak} = 0$), and small dissipation coefficient and modulation depth ($\gamma, \alpha \ll 1$), Eq. (5) becomes

$$2i\frac{\partial \psi}{\partial \tau} = (1-i\gamma)\left(-\nabla^2_{weak} - \mu_1 + |\psi|^2\right)\psi + \alpha_0 \cos(\omega\tau)|\psi|^2\psi, \qquad (6)$$

with $\alpha_0 = \alpha/2$ (for cigar-shaped BECs, $\alpha_0 = \alpha$).

Note that Eq. (6) coincides with Eq. (3) in [1], which was used to predict Faraday patterns in BECs. The wavenumber $k = k(\omega)$ of the excited pattern and the threshold $\alpha_{0,\text{thr}}(k)$ for pattern formation at the first parametric resonance tongue follow from Ref. [1]:

$$k(\omega) = \sqrt{-\mu_1 + \sqrt{\mu_1^2 + \omega^2}}, \qquad \alpha_{0,\text{thr}}(k) = \frac{2\gamma\sqrt{2\mu_1 + k^2}\left(\mu_1 + k^2\right)}{k}. \qquad (7)$$

The behavior predicted by Eqs. (7) when damping, modulation frequency, modulation depth, and reduced chemical potential are all small (the assumptions for validity of Eq. (6)) were confirmed by a direct numerical integration of Eq. (2). We also found that these theoretical predictions qualitatively hold even in weakly dissipative BECs (small $\gamma$) when modulation is stronger and BEC density is larger (with correspondingly larger chemical potential). In order to reduce the computational time, the simulations were carried out on Eq. (2) in 2D space, by assuming a tight (weak) confinement in the vertical (horizontal) direction. Periodic boundary conditions along the horizontal direction were used, corresponding to the limiting case of flat potential in the elongated direction. Figure 1 shows an example of the numerically obtained quasi-1D Faraday patterns. A sequence of snapshots of the BEC density in coordinate space is shown at time intervals equal to half the modulation period (left), and BEC density in momentum space corresponding to snapshot a) is shown on the right. The condensate density pulsates in the tightly confined (vertical) space at the trap modulation frequency $\omega$ [see Eqs. (3) and (4)], whereas the BEC spatiotemporal oscillations along the weakly confined space occur at half the trap modulation frequency. Note that the values of the modulation frequency ($\omega = 0.62$) and of the reduced chemical potential ($\mu_1 = 1.54$) are of order of 1. This means that the mechanism of excitation of Faraday patterns is efficient even when the parameters of the system are beyond the smallness assumptions underlying Eq. (5).

The parametric process usually saturates for sufficiently large dissipation, leading to a stationary pattern (i.e. the pattern steadily oscillating with the half of excitation frequency). However for very weak dissipation the direct parametric process can be followed by its inverse process, where the spatial modulation in weak confinement space converts back into temporal modulation in tight confinement space, which leads to periodical sequences of revivals. We observe periodic revivals in our numerical integration (typically for $\gamma \leq 0.001$), or damped periodic revivals (for the parameter range used in calculation, i.e. for $\gamma \propto 0.01$), however we do not investigate the phenomenon here in detail. Similar periodic revivals in conservative one-dimensional BECs were reported in [11].

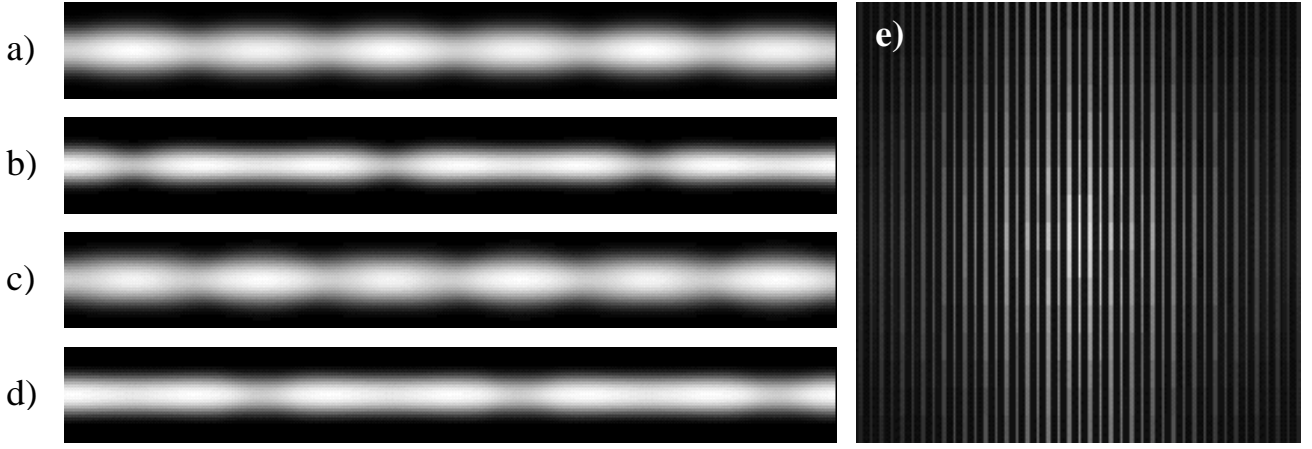

*Figure 1.* a)- d): Sequence of BEC density distributions as taken at every 1/2 of the trap modulation period (from top to bottom); e): BEC momentum distribution (density of the spatial Fourier spectrum of BEC wavefunction) corresponding to snapshot a). Plots are obtained by numerical integration of Eq. (2) with periodic boundary conditions, and with the trapping potential in vertical (Z) direction only. The trap modulation frequency is $\omega = 0.62$ ($\omega_{breath} \approx 1.77$). Other parameters are: $\alpha = 0.5$, $\gamma = 0.01$, $\mu = 1.54$. The spatial grid is 256×32, (aspect ratio: 8:1). The size of integration space along the horizontal (X) coordinate is 176. The mode $n = 3$ of periodic boundaries (along X axis) is excited.

A key aspect of the modulation frequency value used in Fig.1 is that it is far below the natural oscillation frequency of the BEC breathing mode in the tight confinement direction, $\omega_{breath}$. This breathing mode corresponds to periodic variations of the condensate width and peak density at a frequency equal to $\omega_{breath} \approx 2$, i.e. $\Omega_{breath} \approx 2\Omega_{tight}$, and can be efficiently excited through the modulation of the trap potential [7]. In the case of Fig. 1 $\omega_{breath} \approx 1.77$ as obtained numerically by perturbing the undriven BEC in ground state. One can expect that the breathing mode manifests as a resonance in the pattern forming threshold. An example of the obtained dynamics as the modulation frequency increases is shown in Fig.2, which is analogous to Fig.1 but with a modulation frequency ($\omega = 1.54$) closer to expected resonance. A main effect observed is the enhancement of the contrast of the pattern, as well as a dramatic lowering of the threshold for pattern formation. In fact, the thresholds, as found numerically, are $\alpha_{thr} = 0.149$ and $\alpha_{thr} = 0.031$ for Figs.1 and 2, respectively.

In order to quantify this resonance, we have computed numerically the threshold and the selected wavenumber, and compared them with Eqs. (7). A resonance around $\omega \approx 1.77$ can be anticipated as a reason of the lowering of the threshold in Fig.3.a. One should keep in mind that, due to periodic boundaries, the set of spatial (Fourier) modes is discrete, hence the wavenumber has a discrete character too, whose value increases stepwise by continuously increasing the excitation frequency. This explains why the observed pattern formation threshold has a periodic-like character and lowers for some frequencies (those resonant with BEC modes associated with the weakly confined direction). Fig.3.b shows numerically calculated and analytically predicted wavenumbers, also showing a satisfactory agreement.

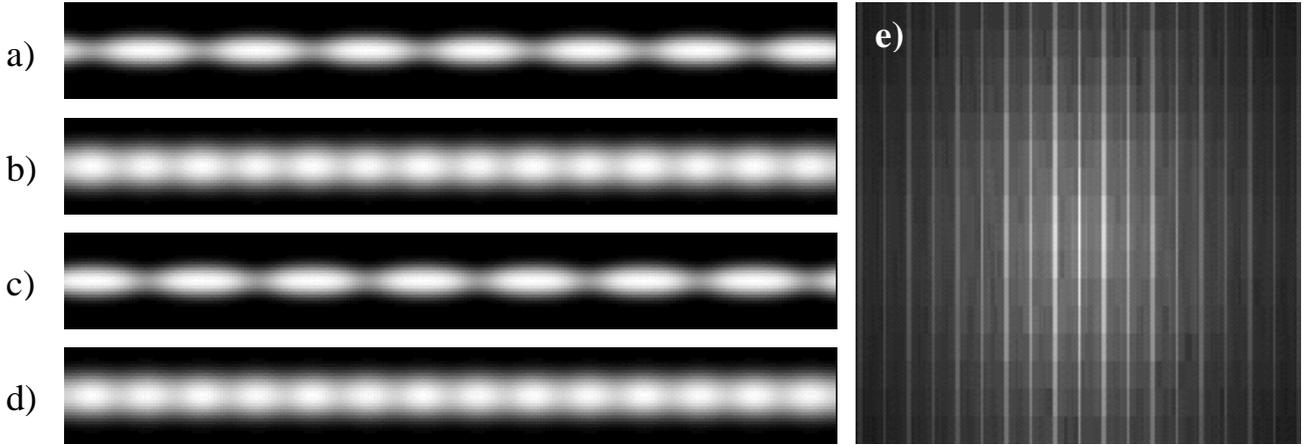

*Figure 2.* Same as Fig. 1, except for parameters: the trap modulation frequency $\omega = 1.56$ is closer to the frequency of the BEC breathing mode in the vertical direction: $\omega_{breath} \approx 1.77$; $\alpha = 0.1$. The mode $n = 6$ of periodic boundaries (along X axis) is excited.

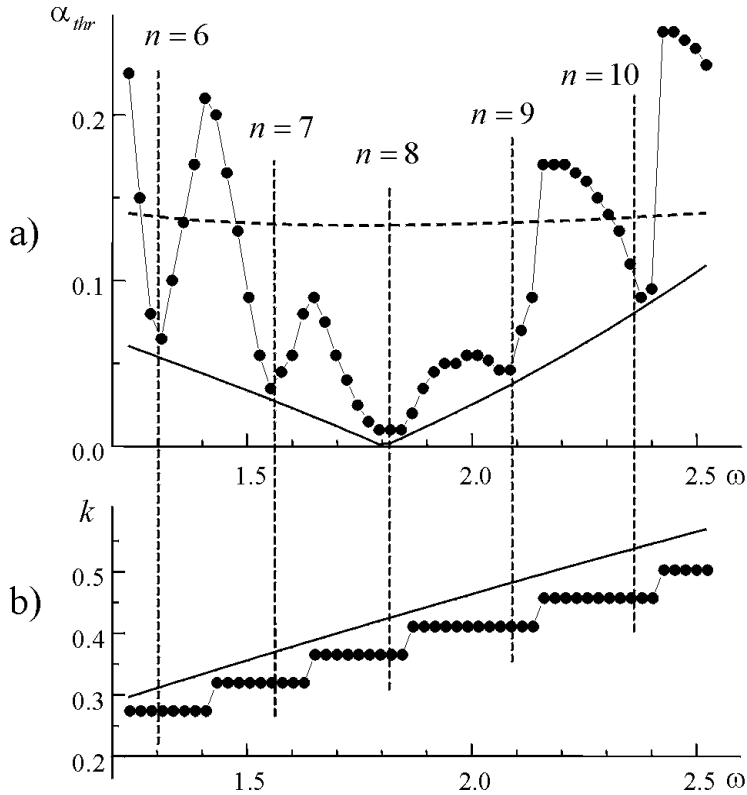

*Figure 3.* a) Pattern formation threshold versus modulation frequency (neutral stability curve) as obtained by numerical integration of Eq. (2) (solid circles), by analytical asymptotic analysis [Eqs. (7), dashed line] and by analytical model taking into account the resonant enhancement of BEC pulsation [Eqs.(8), solid line]. b) Wavenumber of the pattern as obtained by numerical integration of Eq. (2) (solid circles), and by the analytical model (line). Parameters and conditions are the same as in Fig. 1.

When the modulation frequency $\omega$ becomes comparable with $\omega_{breath}$, the description of the BEC dynamics by a lower-dimensional equation can not be performed owing to a strong coupling between tight and weak dimensions. Nevertheless one can clearly relate the resonance in the threshold curve of Fig. 2(a) with the BEC breathing mode along the tight confinement space. In the conservative case, and without modulating the trap frequency, such breathing oscillations are undamped [7], and in an elongated BEC may lead to a self-parametric instability, i.e the transfer of the breathing energy in tight confinement space to longitudinal waves in a weak confinement space [12]. In the presence of damping and modulation of the trap frequency, the breathing oscillations can be parametrically excited, and the amplitude of oscillation for a given modulation depth shows a resonant peak when the parametric resonance condition $\omega = \omega_{breath}$ is attained [7].

One can evaluate the resonant enhancement of the condensate breathing depth versus the modulation frequency by assuming a BEC with large aspect ratio (we set $\omega_{weak} = 0$) and by using a Gaussian-shaped Ansatz to Eq. (2) with time dependent amplitude and width. For definiteness, we consider a disk-shaped BEC and set $u(X,Y,Z,\tau) = A(\tau)\exp[-\beta(\tau)Z^2]$, where $A(\tau)$ and $\beta(\tau)$ are complex-valued functions of time. (Note the assumed independence of $u$ on the weakly confined transverse directions.) After substituting the Ansatz into Eq. (2) and making a parabolic approximation in $Z$ of the nonlinear term one obtains:

$$\frac{dA}{d\tau} = -(\gamma + i)(\beta + |A|^2 - \mu)A \ , \tag{8.a}$$

$$\frac{d\beta}{d\tau} = -(\gamma + i)\left[2\beta^2 + |A|^2(\beta + \beta^*) - \tfrac{1}{2}\omega_{tight}^2(\tau)\right]. \tag{8.b}$$

In the absence of modulation [$\omega_{tight} = 1$, Eq. (3)] the steady-state solution to Eqs. (8), corresponding to the BEC ground state, reads $\overline{\beta} = 1/(4\mu)$, $|\overline{A}|^2 = \mu - 1/(4\mu)$, which imposes $\mu > 1/2$. The response of the condensate to small modulation of the trap frequency can be easily studied in the limit $\gamma, \alpha \ll 1$ by standard linearization of Eqs. (8) around the steady-state solution. The amplitude $\Delta|A|^2$ of the forced oscillation of the condensate peak density around its mean value $|\overline{A}|^2$ can be determined, however leads to not analytically tractable results. Some asymptotic relations are, however, possible in the limit $\gamma \to 0$. In particular for a disk-shaped condensate the resonance frequency is $\omega_{res}^2 = 3 + 1/(4\mu^2)$, which corresponds well with the numerical result $\omega_{breath} \approx 1.77$ for numerical parameters used. (Note that $\omega_{breath} \to 2$ when $\mu \to 1/2$, i.e., when $|\overline{A}|^2 \to 0$, which is the weak nonlinear interaction limit.) The modulation enhancement factor $\delta = \left(\Delta|A|^2/|\overline{A}|^2\right)/\alpha$ reads $\delta = 2/|\omega^2 - \omega_{res}^2|$, and for $\omega \ll \omega_{tight}$ is close to 1/2, i.e. corresponds well to that following from the asymptotic expansion (6): $\alpha_0/\alpha = 1/2$. The analogous calculations for the cigar-shaped condensate leads to $\omega_{res} = 2$, and $\delta = 4/|\omega^2 - \omega_{res}^2|$ independently on the value of $\mu$, which is compatible with [7].

The above analysis concerns the absence of the confinement in the weak confinement direction. Figure 4 shows the numerical calculations in presence of weak confinement, with parameters compatible to experimentally realized elongated BECs (e.g. reported in [13]). The pattern formation thresholds as well as the dominating wavenumbers in the presence of weak confining potentials correspond satisfactorily to those calculated above for the idealized case of the condensates infinitely extended in the weak confinement space. The modulation depth parameter was chosen approximately 2 times above its threshold value, and the pattern, as shown in Fig. 4, appeared after 50 periods of the trap modulation.

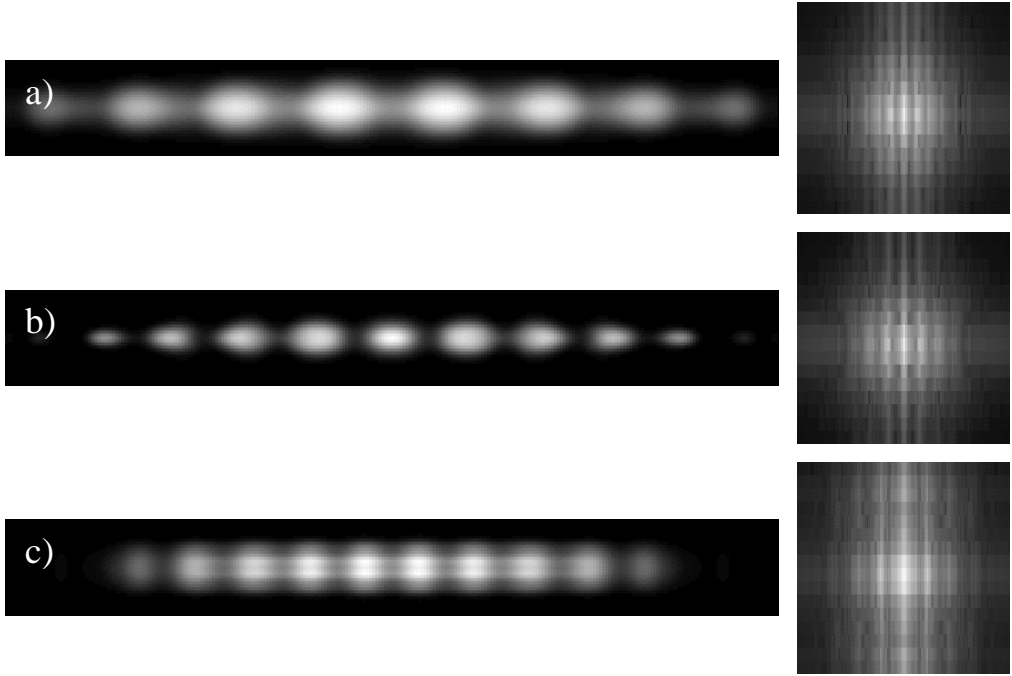

*Figure 4. Snapshots of the BEC density distributions (left: coordinate space; right: momentum space) as obtained by numerical integration of Eq. (2) with weak (nonzero) confinement potential. The aspect ratio of the trap is $\Delta X/\Delta Y \approx (\omega_{tight}/\omega_{weak})^{1/2} = 8$. Parameters and conditions as in Fig. 1, and: a) $\omega = 1.55$, $\alpha = 0.15$; b) $\omega = 2.05$, $\alpha = 0.15$, c) $\omega = 2.95$, $\alpha = 0.75$.*

We note finally that the dissipative patterns, as shown in Figs.1, 2, and 4 are stationary (i.e. steadily oscillating with the half of the excitation frequency, but stationary on large time scales.) The stationarity is due to dissipation, which saturatates the parametric instability. In the conservative cases the parametric driving eventually heats and destroys the condensate. However typically the BEC dynamics during the transients of evolving parametric instability is very similar to the dynamics reported for the dissipative BECs in Figs.1, 2, and 4.

Concluding, we demonstrated that low- (one- and two-) dimensional Faraday patterns can be parametrically excited in the *weak* confinement space of BECs by periodic modulation of the trap frequency in the *tight* confinement space. The reported mechanism is alternative to the modulation of the scattering length studied in [1]. Since the modulation of the trap parameter is often easier than the modulation of the scattering length, and additionally leads to a dramatic threshold reduction due to the excitation of the internal oscillation mode of the condensate, we envisage that our results can have major importance toward an experimental observation of Faraday patterns in BECs.

Financial support from Sonderforschungsbereich 407 of Deutsche Forschungsgemeinschaft (K.S) and from the Spanish Ministerio de Ciencia y Tecnología and European FEDER funds through project BFM2002-04369-C04-01 (G.deV.) is acknowledged. K.S. gratefully aknowledges discussions with M. Lewenstein, L. Santos, J. Arlt, and C.O. Weiss.